# Tunable Topological States in Electron-Doped HTT-Pt

Xiaoming Zhang[1], Mingwen Zhao[*,1], Feng Liu[2,3]

[1] School of Physics and State Key Laboratory of Crystal Materials, Shandong University, Jinan 250100, Shandong, China.

[2] Department of Materials Science and Engineering, University of Utah, Salt Lake City, Utah 84112, USA

[3] Collaborative Innovation Center of Quantum Matter, Beijing 100084, China

**ABSTRACT:** Driving existing materials to exhibit topologically nontrivial state is of both scientific and technological interests. Using first-principle calculations, we propose the first demonstration of electron doping induced multiple quantum phase transition in a single material of the organometallic framework, HTT-Pt, which has been synthesized by reacting triphenylene hexathiol molecules (HTT) with $PtCl_2$. At low electron doping, the HTT-Pt converts from a normal insulator to a quantum spin Hall (QSH) insulator with time-reversal symmetry (TRS). At high electron doping, the TRS is further broken making the HTT-Pt a quantum anomalous Hall (QAH) insulator. The topologically nontrivial band gap of the electron-doped HTT-Pt opened by intrinsic spin-orbit coupling (SOC) can be as large as 44.5 meV, which is promising for realizing these quantum phases at high temperatures. The possibility of switching between the QSH and QAH states offers an intriguing platform for new device paradigm by interfacing between a QSH and QAH state.

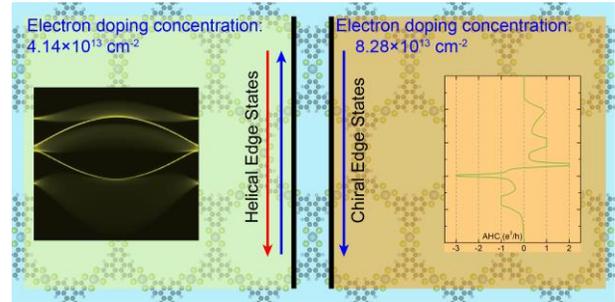

**KEYWORDS:** *Multiple quantum phase transition, electron doping, QSH and QAH insulator, large SOC gaps, QSH/QAH interface*

As the semiconductor industry continually shrinks the size of electronic components on silicon chips, the limits of the current technologies are reached due to the smaller size being prevented by the fundamental physical laws. Spintronics, with the advantage of high density and nonvolatility of data storage, the fast data processing, and low-power-consumption, has the potential to revolutionize the electronic devices. Topological insulators (TIs) provide a promising class of spintronics materials because of their exotic dispersive bands and topologically nontrivial electronic states. TIs have a bulk energy gap, which is linked by gapless edge/surface states that facilitate quantized electronic conduction on the boundaries. The edge states in two dimensional (2D) TIs with time-reversal symmetry (TRS), referred to as quantum spin Hall (QSH) states,[1] can host spin-current with different spin orientations propagate in the helical edge states with opposite directions, where spin-orbit coupling (SOC) plays the role analogous to the external magnetic field in quantum Hall systems. QSH states can also give rise to the so-called quantum anomalous Hall (QAH) states[2] when TRS is broken via inducing magnetism,[3,4] where only one spin orientation propagates in the chiral edge states without need of external magnetic field.

The ability to control electronic properties of a material by applying external electric voltage is crucial for spintronics device applications. The electric field allows one to realizing electron or hole doping and, consequently, change the position of Fermi level. A suitable gate dielectric material can offer both low-temperature and low-voltage processability via increasing capacitance. In conventional solid-state field-effect transistors,[5] the electron doping concentration can reach ~$10^{12}$ $cm^{-2}$, at which the electrons behave as chiral particles accompanied by unexpected physical phenolmena.[6,7] Using a solid polymer electrolyte gate with higher capacitance,[8] the electron doping concentration in bilayer graphene can be up to $5\times10^{13}$ $cm^{-2}$, resulting in a significant band gap and breaking of inversion symmetry.[9,10] In recent experiments, the doping level for both electrons and holes was increased to ~$10^{14}$ $cm^{-2}$ by employing high-capacitance gate insulator,[11] such as Li salt ($LiClO_4$ or Li bis(trifluoromethylsulfonyl)imide (LiTFSI)) dissolved inpoly (ethylene oxide) (PEO),[12] which can be used to modulate magnetoresistance[13] and induce superconductivity at the surface of insulators.[14,15]

Tuning the position of Fermi level by gate voltage is often inevitable for detecting topologically nontrivial states in TIs, because in most cases the Fermi level is not exactly in the bulk band gap.[16-18] For example, the already-synthesized 2D $Ni_3C_{12}S_{12}$,[19] which is predicted to be one of the versions of organic topological insulators (OTIs),[16-18] has the Fermi level locating at the trivial band gap. Electron doping at the level of $2\times10^{14}$ $cm^{-2}$ is needed to move the Fermi level to the topologically nontrivial band gap. Other OTIs, such as $Au_3C_{12}S_{12}$,[19] $Mn_3C_{12}S_{12}$,[20] $Ni_3(C_{18}H_{12}N_6)$,[21] and $Au_3(C_{18}H_{12}N_6)$,[22] also require doping electrons or holes with the doping concentrations magnitude of $10^{13}$ $cm^{-2}$. We note that in these cases the doping mainly tunes the position of the Fermi level, but the electronic band topology of the OTIs remains intact.

In this contribution, we demonstrate theoretically that electron doping can not only regulate the Fermi level but also modify the electronic band topology of OTIs, leading to a stable ferromagnetic ordering along with the topological nontriviality. Specifically we show that the already-synthesized organometallic framework

(OMF), HTT-Pt, can be tuned to a $Z_2$ or a Chern TI by electron doping. At a doping concentration of $4.14 \times 10^{13}$ cm$^{-2}$ (two additional electrons in one unit cell), the HTT-Pt is converted into a QSH insulator, as characterized by a nonzero $Z_2$ topological invariant. When the doping concentration is further increased to $8.28 \times 10^{13}$ cm$^{-2}$ (four additional electrons in one unit cell), spin-polarization takes place spontaneously, giving rise to a QAH state with a finite Chern number of $C=1$ at the Fermi level. The topologically nontrivial bulk band gap due to intrinsic SOC of platinum atoms can be as large as 44.5 meV, corresponding to a temperature of 513.5 K. The possibility of switching between QSH and QAH states in a real material via electrical gating (doping) offers not only an intriguing platform for fundamental study of quantum phase transitions but also new device paradigm by interfacing between a QSH and QAH state.

Our first-principles calculations are performed using Vienna ab initio simulation pack (VASP).[23, 24] For the electron-electron interaction, we employ a generalized gradient approximation (GGA) in the form of Perdew-Burke-Ernzerhof (PBE).[25] The electron-ion interaction is described by projector-augmented-wave (PAW) potentials.[26] The vacuum region is about 15 Å to avoid mirror interactions between neighboring images. Structural optimizations are performed using a conjugate gradient (CG) method until the remanent force on each atom was less than 0.005 eV/Å. A plane-waves energy cutoff of 500 eV is used on a $1 \times 1 \times 1$ Monkhorst-Pack sampling in Brillouin zone (BZ) for CG calculations and on a $3 \times 3 \times 1$ sampling for total energy calculations, respectively. The electron doping effect is studied by adding additional electrons to the lattice with a homogeneous background charge of opposite sign, as done before for other proposed OTIs.[16, 27] Electronic spin-polarization is tested in all calculations.

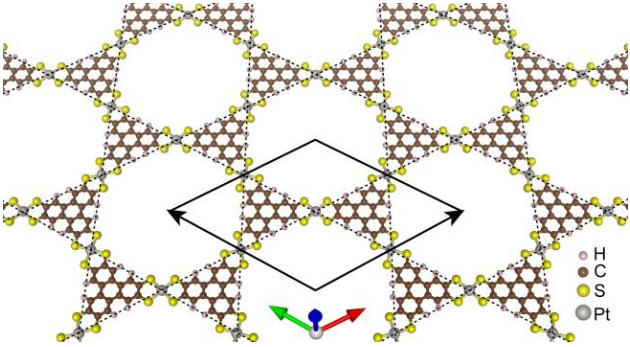

**Figure 1.** Schematic representation of HTT-Pt with the unit cells indicated by the solid lines and the kagome lattice pattern by dashed lines.

A HTT molecule contains a triphenylene core and three chelating dithiolenes. The structural symmetry of the core and the feasibility of covalent metal-dithiolene links make it an ideal molecular building block in the syntheses of porous polymer OMFs. In experiments, the Pt-contained OMF, HTT-Pt was synthesized by directly reacting HTT with PtCl$_2$.[28] Each HTT coordinates with three Pt atoms, forming a structurally perfect kagome lattice pattern as sketched by dashed lines in Figure 1. After structural optimization, the planar configuration and six-fold symmetry are both well preserved, which can be attributed to the fully conjugate features of the framework. The optimized equilibrium lattice constant of the HTT-Pt monolayer is $a = 23.61$ Å, slightly longer than the experimental value (23.29 Å) measured from a staggered stacking among the neighboring HTT-Pt sheets, whereas the length of Pt-S distance (2.27Å) is shorter than the experimental value (2.36 Å). The lengths of the C-S and C-H bonds are 1.72 Å and 1.09 Å, respectively. The covalent bonds between carbon atoms vary from 1.40 to 1.46 Å in length, very close to that in graphene (1.42 Å).

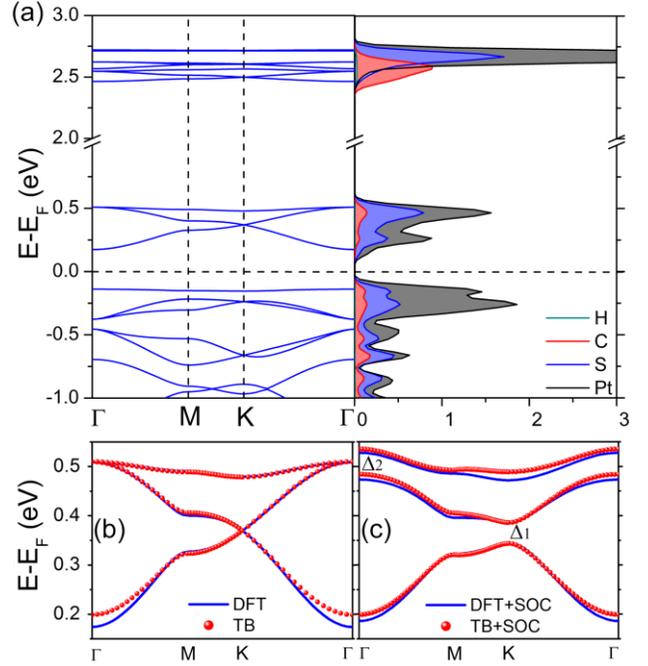

**Figure 2.** (a) Band structure and projected density of states (PDOS) of HTT-Pt without SOC. The energy at the Fermi level was set to zero. (b, c) A comparison between the band structures calculated from first-principles (DFT) and TB models (b) without and (c) with SOC.

Our DFT calculations indicated that the pristine HTT-Pt is a normal insulator with a trivial band gap of 0.313 eV at the $\varGamma$ point (see the left panel of Figure 2a). There are typical kagome bands, composing one flat band above two Dirac bands in the conduction band region. The electronic density of states (PDOS) projected onto different atoms are shown in the right panel of Figure 2a. We can clearly see the kagome bands come mainly from the platinum and sulphur with minimal contribution from carbon and none contribution from hydrogen atoms. This property is further confirmed by the partial charge density isosurfaces decomposed to the kagome bands, as shown in Figure 3a. An ideal kagome lattice pattern can be extracted out from the spatial distribution of the isosurfaces, which allows us to reproduce the kagome bands by using a single-orbital TB model involving the nearest-neighbor (NN) and next-nearest-neighbor (NNN) interactions.[29] The Hamiltonian in the reciprocal space can be expressed as follows:

$$H_0 = \begin{pmatrix} \varepsilon_{on} & 2t_1 \cos k_1 & 2t_1 \cos k_2 \\ 2t_1 \cos k_1 & \varepsilon_{on} & 2t_1 \cos k_3 \\ 2t_1 \cos k_2 & 2t_1 \cos k_3 & \varepsilon_{on} \end{pmatrix}$$
$$+ \begin{pmatrix} 0 & 2t_2 \cos(k_2+k_3) & 2t_2 \cos(k_3-k_1) \\ 2t_2 \cos(k_2+k_3) & 0 & 2t_2 \cos(k_1+k_2) \\ 2t_2 \cos(k_3+k_1) & 2t_2 \cos(k_1+k_2) & 0 \end{pmatrix} \quad (1)$$

where $k_1 = k_x a/2, k_2 = k_x a/4 + \sqrt{3} k_y a/4$ and $k_3 = -k_x a/4 + \sqrt{3} k_y a/4$. $\varepsilon_{on}$ is the on-site energy. $t_1$ and $t_2$ correspond to the NN and NNN hopping parameters (see Figure 3a), respectively. The above TB Hamiltonian can well reproduce the kagome bands obtained from first-principles calculations with the optimal fitting parameters of $\varepsilon_{on} = 406.0$ meV, $t_1 = -46.6$ meV and $t_2 = -5.2$ meV, as shown in Figure 2b.

Taking the SOC into account, our first-principles calculations demonstrate that the degeneracies of the two Dirac bands at $K$ point, and of the upper Dirac band and flat band at $\Gamma$ point are lifted, leading to energy gaps of $\Delta_1 = 42.6$ and $\Delta_2 = 54.5$ meV respectively, as shown in Figure 2c. We attribute such large SOC gap to the strong SOC strength of heavy Pt atom. Similar features have also been predicted in trans-Pt-(NH)S framework.[30] The band gaps can be reproduced by TB model via introducing NN and NNN SOC terms into the Eq. (1). The TB Hamiltonian involving intrinsic SOC reads as:

$$H_{SOC} = H_0 + H_{\pm} \quad (2)$$

$$H_{\pm} = \pm i2\lambda_1 \begin{pmatrix} 0 & \cos k_1 & -\cos k_2 \\ -\cos k_1 & 0 & \cos k_3 \\ \cos k_2 & -\cos k_3 & 0 \end{pmatrix}$$

$$\pm i2\lambda_2 \begin{pmatrix} 0 & -\cos(k_2+k_3) & \cos(k_3-k_1) \\ \cos(k_2+k_3) & 0 & -\cos(k_1+k_2) \\ -\cos(k_3+k_1) & \cos(k_1+k_2) & 0 \end{pmatrix} \quad (3)$$

here $\lambda_1$ and $\lambda_2$ correspond to the NN and NNN SOC parameters (see Figure 3a), respectively, and +/- refers to the spin-up/spin-down channels. The TB bands with the SOC parameters of $\lambda_1 = 6.0$ meV and $\lambda_2 = 1.6$ meV are shown by the red-dotted lines in Figure 2c. This confirms that the energy gaps in HTT-Pt lattice is opened mainly due to the intrinsic SOC of platinum atoms.

To achieve the QSH states, the Fermi level should be moved upward to the kagome band region. This can be experimentally realized by electron doping. Our first-principles calculations indicate that at the doping concentration of $4.14 \times 10^{13}$ cm$^{-2}$ (two additional electrons in one unit cell), the kagome bands remain intact and the Fermi level exactly locates at the Dirac point, as shown in Figure 3b. Such doping concentration is experimentally accessible using the current techniques.[11] The SOC opens a band gap of 44.5 meV at the Fermi level. The topological aspect of the band gap can be determined using the parity criteria proposed by Fu and Kane.[31] In this strategy, the $Z_2$ topological index $\nu$ is defined as

$$(-1)^{\nu} = \prod_i \delta_i \text{ with } \delta_i = \prod_{m=1}^{N} \xi_{2m}(\Gamma_i) \quad (4)$$

for 2N occupied bands. $\xi_{2m}(\Gamma_i) = \pm 1$ is the parity eigenvalue of the $2m$-th occupied energy band at the time-reversal invariant momentum $\Gamma_i$. This method has been widely used to confirm the topological aspects of TIs.[32, 33] The calculated $\delta_i$ of the HTT-Pt are (+), (-), (-), and (-) at the four time-reversal momenta (0, 0), (1/2, 0), (0, 1/2), and (-1/2, 1/2), as shown in Figure 3c, leading to a nonzero $Z_2$ topological invariant of $\nu =1$. This implies the topological nontriviality of the electron-doped HTT-Pt. Since the existence of topological edge states is an important signature of the 2D TIs, we calculated the edge states of the HTT-Pt lattice by using the Wannier90 package.[34] Using the recursive method,[35] the edge Green's function of the semi-infinite lattice was constructed from maximally localized Wannier functions (MLWFs), and the local density of states (LDOS) of the edge was calculated and shown in Figure 3d, where one can see the nontrivial topological edge states that connect the bulk states. It is interesting to see that this SOC gap is well separated from other bands and thus implementable for achieving the QSH effect at room temperature.

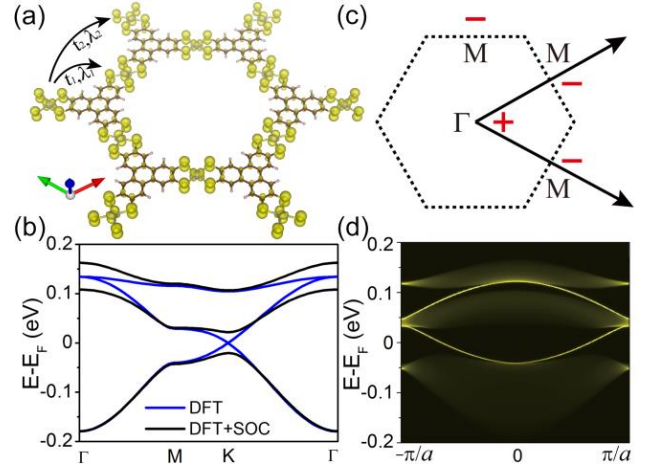

**Figure 3.** (a) The partial charge density isosurfaces decomposed to the kagome bands with the isosurface value of 0.002 Å$^{-3}$. The NN and NNN interactions of the single-orbital TB model are marked in it. (b) Enlarge view of the kagome bands with and without SOC when the Fermi level is moved to the Dirac point via doping two electrons per unit cell. (c) The products of the parities ($\delta$) of the occupied bands at the time-reversal invariant momentum. (d) The semi-infinite Dirac edge states within the SOC gaps.

When the doping concentration is further increased to $8.28 \times 10^{13}$ cm$^{-2}$ (doping four electrons in one unit cell), the Fermi level is expected to be moved into the region between the flat band and the upper Dirac band, provided that the kagome bands remain intact. However, our DFT calculations showed that the TRS of the system was broken at this electron doping concentration, leading to a spin-polarized ground state with a magnetic moment of 2.0 $\mu_B$ per unit cell. The electron spin-polarization is related to instability of the partially-filled flat band. Due to the overlap between flat band and the Dirac bands, as shown in Figure 2b, electrons come to occupy the flat band as the Fermi level is moved upward to this region. The instability of the partially-filled flat band leads to spontaneous electron spin-polarization to reduce energy of the system.[18, 36] As a consequence, the spin degeneracy of the kagome bands is lifted. The kagome bands of one spin channel (up spin) are fully occupied by electrons, while only one Dirac band of another channel (down spin) is filled. The Fermi level is right at the Dirac point of the spin-down channel, as shown in Figure 4a.

To reveal the mechanism of this quantum phase transition, we introduced an exchange field ($M$)[17] into Eq. (1) to describe the spin-polarized kagome bands. The spin-polarized Hamiltonian reads as:

$$H_{spin} = \begin{pmatrix} 1 & 0 \\ 0 & 1 \end{pmatrix} \otimes H_0 - M \sum_{i,\alpha,\beta} c_{i\alpha}^+ s_{\alpha\beta}^z c_{i\beta} \quad (5)$$

here, $c_{i\alpha}^+$ and $c_{i\alpha}$ are creation and annihilation operators, respectively, for an electron with spin α onsite $i$. Diagonalizing the above Hamiltonian in reciprocal space, we obtain the band structures shown in Figure 4a, which reproduce well the spin-polarized kagome bands given by DFT calculations with the optimal parameters of $\varepsilon_{on} = -50.0$ meV, $t_1 = -42.0$ meV, $t_2 = -4.7$ meV, and $M = -83.0$ meV, respectively.

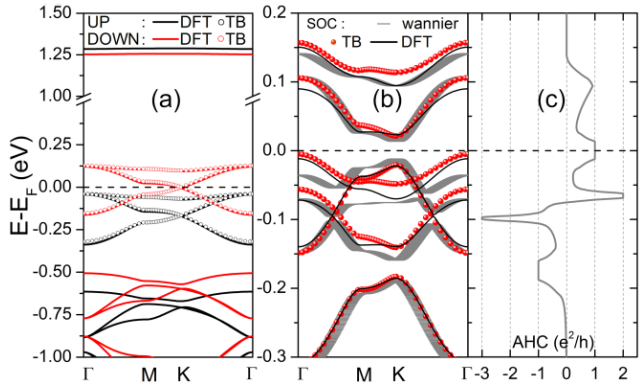

**Figure 4.** (a, b) A comparison between the band structures of the heavy electron-doped HTT-Pt calculated from first-principles (DFT), spin-polarized TB models and MLWFs (a) without and (b) with SOC. (c) The AHC within the energy window of the spin-polarized kagome bands.

Taking SOC into account, our DFT calculations showed that a band gap of $\Delta_3$=42.5 meV is opened at the Fermi level, as shown in Figure 4b. The SOC gap can be reproduced by introducing intrinsic SOC terms $H_+$ and $H_-$ into the Eq. (5) with the SOC parameters of $\lambda_1 = 6.0$ meV and $\lambda_2 = 1.6$ meV (see Figure 4b). The topological nontriviality of the heavy electron-doped HTT-Pt can be evidenced by the Chern number (*C*) of the spin polarized bands calculated using the Kubo formula.[37, 38] Similar strategy has been successfully adopted in confirming the QAH states in a well-designed graphene nanomesh.[39] The calculated anomalous Hall conductivity (AHC) as a function of the electron filling is presented in Figure 4c, which shows the AHC develops plateaus at value of $e^2/\hbar$ at the Fermi level, indicating that the HTT-Pt doped with four electrons per unit cell is a Chern TI characterized by a finite Chern number (*C*=1). The electrical gating (doping) induced quantum phase transition provides the possibility of achieving QSH/QAH interface in a real material.[40]

Notably, compared with the large bulk band gaps predicted in inorganic TIs (0.74~1.08 eV),[41, 42] the bulk gaps of OTIs are usually very small, 2.4~22.7 meV, partly due to the weak SOC of the transition metal atoms. The incorporation of Pt atoms into the trans-Pt-(NH)S framework has been shown to improve the SOC gap to ~50 meV of the QSH state.[30] Similarly, large SOC gaps are predicted here in a real experimental sample, HTT-Pt, and especially the 42.5 meV gap for QAH state is the largest known in an organic system to date. Additionally, breaking TRS in $Z_2$ TIs was commonly achieved by doping with magnetic atoms[3, 43] or introducing proximity coupling with an antiferromagnetic insulator.[44, 45] Our work offers an alternative approach to induce ferromagnetism in TIs that is reversible.

In summary, we demonstrate from first-principles that the already-synthesized HTT-Pt can undergo reversible quantum phase transitions through electron doping, to exhibit tunable topological properties. At low electron doping concentration of $4.14 \times 10^{13}$ cm$^{-2}$, the HTT-Pt coverts from a normal insulator to a 2D TI characterized by a nonzero $Z_2$ topological invariant. As the electronic doping concentration being further increased to $8.28 \times 10^{13}$ cm$^{-2}$, TRS is broken, leading to a robust QAH state with the Chern number of *C*=1. Also, the topologically nontrivial bulk band gap is as large as 44.5 meV (~513.5 K), implying that the critical temperatures of realizing QSH and QAH effects are above room temperature. The possibility of switching between two QSH and QAH states in a real material via electrical gating (doping) offers not only an intriguing platform for fundamental study of quantum phase transitions but also new device paradigm by interfacing between a QSH and QAH state.

## AUTHOR INFORMATION

### Corresponding Author
*E-mail: zmw@sdu.edu.cn.

### Author Contributions
The manuscript was written through contributions of all authors. All authors have given approval to the final version of the manuscript.

### Notes
The authors declare no competing financial interests.

## ACKNOWLEDGMENT

This work is supported by the National Basic Research Program of China (No.2012CB932302), the National Natural Science Foundation of China (Nos.91221101, 21433006), the 111 project (No. B13029), the Taishan Scholar Program of Shandong, and the National Super Computing Centre in Jinan. F.L. acknowledges support from DOE-BES (Grant # DE-FG02-04ER46148).